\documentclass[amsmath,amssymb,aps,prc,twocolumn,superscriptaddress,showpacs,preprintnumbers]{revtex4}
\usepackage{graphicx,color}
\usepackage{multirow}
\usepackage{times}
\usepackage{bm}
\usepackage{epstopdf}
\newcommand{\GeV}{\mathrm{GeV}}
\usepackage[normalem]{ulem}  % \sout{old text} for strikeout

\newcommand{\comment}[1]{}
\renewcommand\sout{\bgroup \color{red} \ULdepth=-.5ex \ULset}
\newcommand{\srtNN}{\sqrt{s_{{\scriptscriptstyle NN}}}}

\begin{document}
\preprint{YITP-15-120}

\title{
Examination of directed flow as a signature of the softest point of
the equation of state in QCD matter
}

\author{Yasushi Nara}
%\email[]{tsutsui@yukawa.kyoto-u.ac.jp}
%\thanks{}
\affiliation{
Akita International University, Yuwa, Akita-city 010-1292, Japan}
\affiliation{Frankfurt Institute for Advanced Studies, 
D-60438 Frankfurt am Main, Germany}
\author{Harri Niemi}
\affiliation{Institut f\"ur Theoretishe Physik,
 Johann Wolfgang Goethe Universit\"at, D-60438 Frankfurt am Main, Germany}
%%%%%
%
\author{Akira Ohnishi}
\affiliation{Yukawa Institute for Theoretical Physics, Kyoto University,
Kyoto 606-8502, Japan}
%%%%%
\author{Horst St\"ocker}
\affiliation{Frankfurt Institute for Advanced Studies, 
D-60438 Frankfurt am Main, Germany}
\affiliation{Institut f\"ur Theoretishe Physik,
 Johann Wolfgang Goethe Universit\"at, D-60438 Frankfurt am Main, Germany}
\affiliation{GSI Helmholtzzentrum f\"ur Schwerionenforschung GmbH, D-64291
Darmstadt, Germany}

\date{\today}

\begin{abstract}
We analyze the directed flow of protons and pions
in high-energy heavy-ion collisions
in the incident energy range
from $\sqrt{s_{{\scriptscriptstyle NN}}}=7.7$ to 27 GeV
within a microscopic transport model.
Standard hadronic transport approaches do not describe
the collapse of directed flow below
$\sqrt{s_{{\scriptscriptstyle NN}}}\simeq 20$ GeV.
By contrast,
a model that simulates effects of a softening of
the equation of state describes well the behavior of directed flow
data recently obtained by the STAR Collaboration~\cite{STARv1}.
We give a detailed analysis of how directed flow is generated.
Particularly, we found that softening of the effective equation of state
at the overlapping region of two nuclei, i.e. the reaction stages
where the system reaches high baryon density state,
is needed to explain the observed collapse of proton directed flow
within a hadronic transport approach.
\end{abstract}

\maketitle

\section{Introduction}

Detecting the QCD phase transition is of primary interest
in current nuclear and particle physics.
Experiments have shown that a new form of strongly interacting matter
is created in high-energy nuclear collisions
at the Relativistic Heavy Ion Collider (RHIC) top energy
and the Large Hadron Collider (LHC) energies~\cite{RHIC2005}.
Theoretical arguments and experimental signals imply 
that this matter is a quark-gluon plasma (QGP)
created well above a transition temperature
at almost zero baryon chemical potential.
The lattice QCD calculations have confirmed that
the transition from hadronic matter to QGP
is a crossover at zero baryon density~\cite{Aoki:2006we}.
 
The next challenge is to discover
the first and/or second-order
phase transition of QCD matter.
Several effective model and approximate calculations of QCD
suggest the existence of the first-order phase boundary
and the critical point
at finite chemical potentials~\cite{PhaseDiagram}.
Various observables, such as collective flows, particle ratios, 
moment of the distributions of conserved charges,
have been measured at various incident energies 
to find evidences of a phase transition and
a critical point~\cite{Aggarwal:2010cw,Kumar:2013cqa}.
Particularly, we shall focus on
the observation of collapse of directed flow
by the STAR Collaboration~\cite{STARv1},
which might be a signal of a first-order phase transition
at high baryon density region
between hadronic matter and quark gluon plasma~\cite{Rischke:1995pe}.

The collective transverse flow~\cite{Hofmann:1976dy,Stoecker:1986ci}
has been utilized to explore the properties of hot and dense matter,
since it reflects the properties of the equation of state (EoS)
in the early stages
of nuclear collisions~\cite{Stoecker:2004qu,Baumgardt:1975qv}.
The existence of bounce off, one of the collective flows,
in heavy-ion collisions
was first suggested in hydrodynamics
~\cite{Hofmann:1976dy,Stoecker:1986ci,Stoecker:2004qu,Baumgardt:1975qv,Flow1}
and examined in experiments at Bevalac~\cite{Bevalac}
at Lawrence Berkeley National Laboratory.
Later, the presence of the softest point near the phase transition
in the EoS~\cite{Hung:1994eq,Rischke:1995pe}
was discussed as a signal for the first-order phase transition,
where the softest point is a local minimum of
the ratio of pressure to energy density $p/\epsilon$
as a function of energy density, leading to a small sound velocity
defined by $v_s^2=dp/d\epsilon$.
In Ref.~\cite{Ivanov:2000dr},
baryon density dependence of the ratio $p/\epsilon$
was investigated within a quasi-particle model.
The softest point of the crossover EoS at vanishing
chemical potential is not very pronounced, but it is predicted
that the EoS with a first order phase transition
exhibits a very pronounced softest point
at large chemical potentials.

Considerations of heavy ion collisions at RHIC energy have raised the question
whether the physics of the often speculated
first-order phase transition or the ``softest point'' should
better be studied at moderate
energies~\cite{Brachmann:1999xt,Stoecker:2004qu} in which
the collapse of the directed flow at $\srtNN=$7-9 GeV was speculated.
Presumably, the EoS of baryon rich matter is
the softest at moderate energy densities of a few GeV/fm$^3$.

Particularly,
the excitation function of the directed flow slope
with respect to rapidity $dv_1/dy$
decreases at $E_\mathrm{inc}>2 A~\GeV$~\cite{Liu:2000am}
and is predicted to exhibit a minimum at a certain collision energy
in hydrodynamical calculations using an EoS with 
a first-order QCD phase transition
~\cite{Rischke:1995pe,Brachmann:1999xt,Ivanov:2000dr},
where the $v_1$ is defined as the first Fourier component in
the azimuthal angle distribution with respect to the reaction plane,
$v_1 = \langle \cos \phi\rangle$.
The negative slope of $v_1$, called
the third flow~\cite{Csernai:1999nf,Csernai:2004gk}
or the anti-flow of the nucleon~\cite{Brachmann:1999xt,Brachmann:1999mp}
emerges as a consequence of 
a tilted ellipsoid with respect to the beam axis
from which negative flow builds up if matter passes through
the softening point.
Thus the collapse of the directed flow slope to a negative value
might signal the first-order phase transition
from hadron phase to quark-gluon phase,
and was recently observed in the beam energy scan (BES) program
performed at RHIC~\cite{STARv1,Shanmuganathan:2015qxb}.

A negative slope of proton $v_1$ at midrapidity has been found in 
the microscopic transport models RQMD~\cite{Snellings:1999bt},
UrQMD~\cite{Bleicher:2000sx},
and PHSD/HSD~\cite{Konchakovski:2014gda}
in which
sign change is purely geometrical and only happens at large impact
parameters and sufficiently higher collision energies.
Note that such microscopic transport models do not show a
negative slope of the proton $v_1$ at midrapidity at bombarding energies
of $\srtNN\leq 20$ GeV
~\cite{Konchakovski:2014gda,Petersen:2006vm,Steinheimer:2014pfa},
and thus the negative slope of the proton $v_1$ at 
$\srtNN=11.5$ and 19.6 GeV observed
by the STAR Collaboration~\cite{STARv1}
is incompatible with the predictions by the standard hadronic
transport models.
We also note that the hadronic transport model
with momentum dependent mean field
significantly improves the description of the directed flow
data from E895~\cite{Liu:2000am} and NA49 data~\cite{NA49prl,NA49prc},
but inclusion of the hadronic mean field does not
lead to the negative proton $dv_1/dy$~\cite{Isse}.
This gives indirect evidence of the phase transition
around such collision energies.

The excitation function of the directed flow slope was
investigated also in
a transport + hydrodynamics hybrid approach~\cite{Steinheimer:2014pfa},
where it was
found that there is no sensitivity of the directed flow on the EoS,
and there is no minimum in the excitation function
of the directed flow slope.
In contrast,
strong sensitivities of the directed flow to the EoS are found
in a three-fluid model~\cite{Ivanov:2014ioa}.
The three-fluid calculations indicate that the crossover deconfinement
transition is consistent with the directed flow data
of energy range up to $\srtNN\approx 11.5$ GeV.
However, the PHSD transport model which incorporates crossover EoS
does not show the experimentally observed minimum~\cite{Konchakovski:2014gda}.
Thus it is not yet clear whether the negative slope of $v_1$
signals the softening of the EoS in hybrid approaches.

In the present paper, 
we investigate the directed flow in the BES energy region
within the microscopic transport model JAM~\cite{JAM} by imposing
attractive orbits for each two-body scattering
to simulate effects of a softening of the EoS.

\section{Model}

The hadronic transport model JAM~\cite{JAM}
has been developed
based on resonance and string production and their decay,
which is similar to other transport
models~\cite{RQMD1995,UrQMD1,UrQMD2}.
Secondary products from decay can interact with each other by binary collisions.
A detailed description of the JAM model can be found in Ref.~\cite{JAM}.

We take into account nuclear EoS effects
within a microscopic transport approach
by changing the standard stochastic two-body scattering style,
which is normally implemented so as not to contribute to the pressure.
For example, we can
simulate repulsive $NN$ potential effects by allowing
only repulsive orbits in the two-body
collisions~\cite{Halbert:1981zz,Gyulassy:1981nq,Kahana:1994be},
instead of choosing the scattering angle randomly
as in a standard cascade.
It is reported that directed flow at Bevalac and Alternating Gradient
Synchrotron (AGS) energies are well
described by this approach~\cite{Kahana:1994be}.
Later, attractive orbits were introduced to effectively incorporate
the softening of the EoS~\cite{Sorge:1998mk}
guided by the virial theorem~\cite{Danielewicz:1995ay}.
In this way, different treatment of scatterings can modify the EoS.

We impose attractive orbits for each two-body hadron-hadron scattering
to reduce the pressure of the system.
The pressure of a system, in which particles are interacting with
each other only by two-body scattering,
is given by the virial theorem as~\cite{Danielewicz:1995ay}
\begin{equation}
 P =  P_f + \frac{1}{3TV} 
  \sum_{(i,j)} 
\left(\bm{q}_i\cdot\bm{r}_i + \bm{q}_j\cdot\bm{r}_j \right)
\end{equation}
where $P_f=1/(3VT)\int dt\sum_i \bm{p}_i\cdot\bm{v}_i$
 corresponds to the free streaming contribution.
The second term represents the pressure generation
from all two-body scatterings between the pair of particles $i$ and $j$,
where $\bm{q}_{i}=-\bm{q}_j=\bm{p}'_i-\bm{p}_i$
is the momentum transfer
and $\bm{r}_i$ and $\bm{r}_j$ are the coordinate
of colliding particles.
$V$ is the volume of the system, and $T$ is a time interval
over which the system is measured.
Thus,
pressure generation by the two-body collisions is related to
the scattering style;
the repulsive orbit $\bm{q}_{i}\cdot(\bm{r}_{i}-\bm{r}_j)>0$
enhances the pressure,
while the attractive orbit $\bm{q}_{i}\cdot(\bm{r}_{i}-\bm{r}_j)<0$
reduces the pressure.
Note that 
an attractive potential softens the EoS~\cite{Li:1998ze},
which leads to an attractive orbit.

Attractive orbits are realized in the simulation as follows.
Each orbit is selected randomly as in the standard simulation,
but in the case where the orbit is repulsive, we change it to an attractive one
by exchanging the momentum of two particles in the two-body cener-of-mass
(c.m.) frame.
Thus the scattering rate remains the same.
While in reality modification of the scattering style should depend
for example, on variables such as local energy density,
we impose a modified scattering style for all hadron-hadron
$2\to2$ scatterings in order to see an effect of the softening,
instead of trying to fit the data.
Thus there is no adjustable free parameter in our current approach
unlike Ref.~\cite{Kahana:1994be} in which
repulsive trajectories are selected for colliding
baryons with some probability in order to generate more pressure
at AGS energies.

An energy density dependent implementation of attractive orbits
will be examined shortly in Sec.~\ref{sec:eosdep}

\section{Results}

We now discuss directed flows in the BES energy region.
In the simulation, we choose the impact parameter range
$4.6<b<9.4$ fm for mid-central 
and $0<b<4.0$ fm for central collisions
for the STAR data~\cite{STARv1}.

\subsection{Beam energy dependence of directed flow}
%
%%%%%%%%%%%%%%%%%%%%%%%%%%%%%%%%%%%%%%%%%%%%%%%%%%%%%%%%%
\begin{figure}[tbh]
\includegraphics[width=9.0cm]{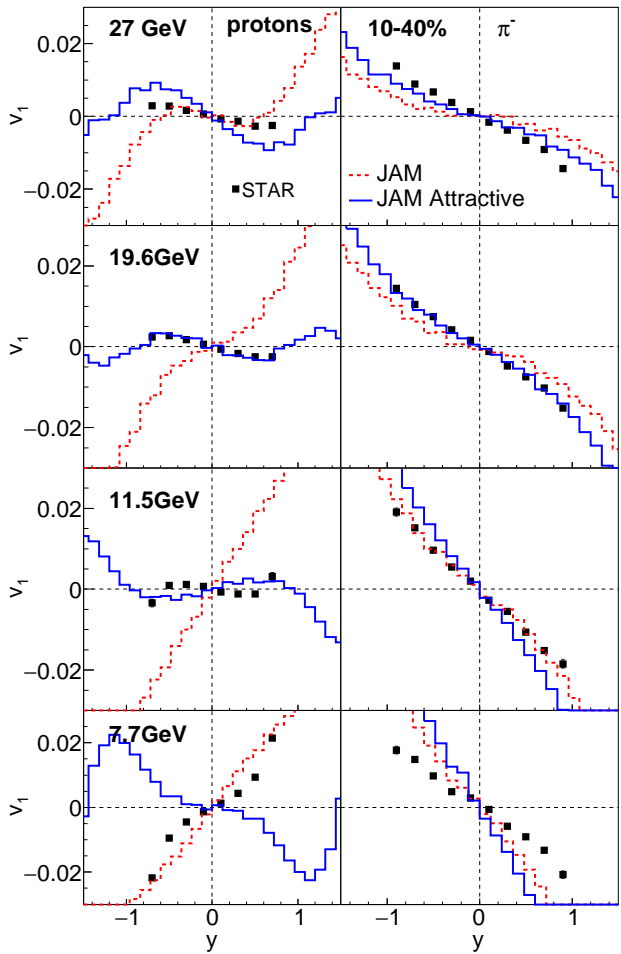}
\caption{Directed flows of
protons and pions in mid-central Au+Au collisions (10-40\%) at 
$\srtNN=7.7$-27 GeV from the JAM cascade mode (dashed lines)
and the JAM cascade with attractive orbits (solid lines)
in comparison with the STAR data~\cite{STARv1}.}
\label{fig:starv1mid}
\end{figure}
%%%%%%%%%%%%%%%%%%%%%%%%%%%%%%%%%%%%%%%%%%%%%%%%%%%%%%%%%
%
In Fig.~\ref{fig:starv1mid}
we show the calculated directed flow $v_1$
of protons and pions in mid-central collisions from
the standard JAM cascade (dotted lines) and
the JAM cascade with attractive orbits (solid line)
in Au+Au collisions at $\srtNN=7.7, 11.5, 19.6$ and 27 GeV
in comparison with the data from STAR Collaboration~\cite{STARv1}.
The standard JAM cascade calculation agrees with the 7.7 GeV data.
However, it is seen that $v_1$ from the standard JAM cascade calculations
for beam energies of 11.5 and 19.6 GeV yields
much larger $v_1$ than the STAR data.
The proton slope from JAM turns out to be negative at $\srtNN=27$ GeV.
This is because of the geometrical reason 
pointed out in Ref.~\cite{Snellings:1999bt},
and is not related to the softening of the EoS
within our transport approach.
We note that our results are consistent with other microscopic transport
approaches~\cite{Konchakovski:2014gda,Petersen:2006vm,Steinheimer:2014pfa}.

By comparison,
attractive orbit scatterings
drastically reduce
the $v_1$ slope, and explain the STAR data
at $\srtNN\gtrsim 10~\GeV$ as shown
in Fig.~\ref{fig:starv1mid} (solid lines);
at $\srtNN=11.5$ and $19.6$ GeV,
the $v_1$ slope becomes almost zero and negative, respectively.
At lower energy $\srtNN=7.7~\GeV$, results with attractive orbits
are far from the data, and there should not be large EoS softening.

From this analysis, we find that the softening of the EoS affects
the directed flow of protons at midrapidity and should emerge
in the beam energy range of
$\srtNN\gtrsim 10~\GeV$, but its effects should be small at $\srtNN=7.7~\GeV$.
Since NA49 data at $\srtNN=8.9$ GeV may also indicate
evidence of softening of the EoS~\cite{Petersen:2006vm},
the onset beam energy of the softening might be lower than 10 GeV.
Therefore, detailed experimental studies are needed 
around the beam energies of $\srtNN\lesssim 10$ GeV.

Unfortunately, the EoS softening effects are not easy to see
when the $v_1$ slope is already negative in the standard cascade.
As seen in Fig.~\ref{fig:starv1mid},
the proton $v_1$ slope at $\srtNN=27~\GeV$ and pion $v_1$ slopes
are negative in the standard cascade
from geometrical non-QGP effects~\cite{Snellings:1999bt}
and from absorption by baryons~\cite{Bass:1995pj}, respectively.
It should be noted,
however, that JAM with attractive orbits overestimates
the negative slope of the proton $v_1$ indicating the need to
reharden the EoS,
i.e.; matter created at this collision energy reaches
well above the transition region or weak softening
of the EoS due to less net baryonic density.

%%%%%%%%%%%%%%%%%%%%%%%%%%%%%%%%%%%%%%%%%%%%%%%%%%%%%%%%%
\begin{figure}[tbh]
\includegraphics[width=9.0cm]{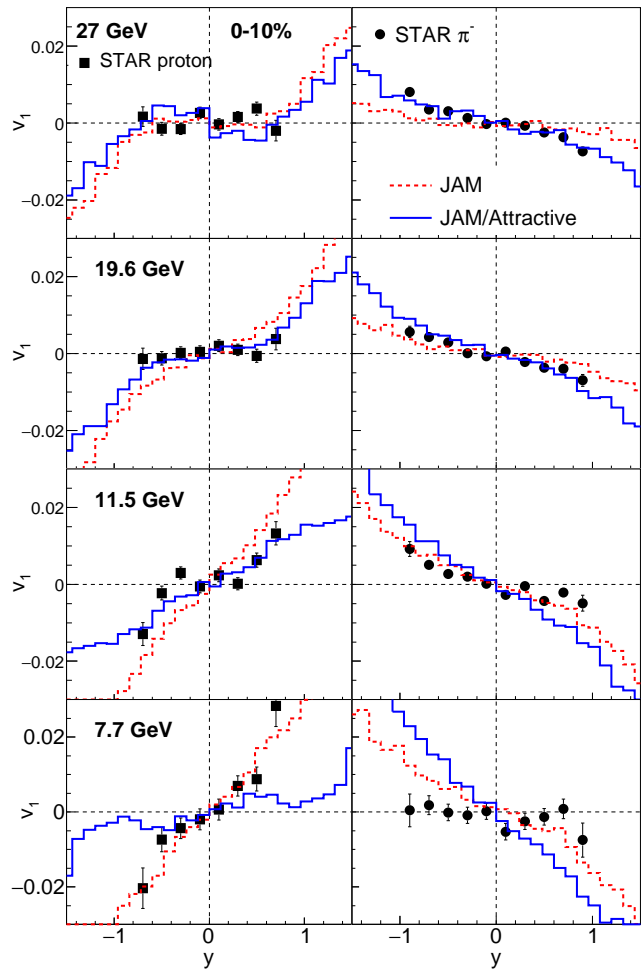}
\caption{Same as in Fig.~\ref{fig:starv1mid}, but for
central collisions (0-10\%).
}
\label{fig:starv1cent}
\end{figure}
%%%%%%%%%%%%%%%%%%%%%%%%%%%%%%%%%%%%%%%%%%%%%%%%%%%%%%%%%

We now discuss the directed flows in central collisions
in order to distinguish the geometrical effects from phase transition.
Since sign change of the proton $v_1$ is purely geometrical
and only happens at large impact parameters in standard
hadronic transport models, 
it is possible to find the effects of the softening
sharply in central collisions.
We show directed flows in central collisions in Fig.~\ref{fig:starv1cent}.
STAR data on the proton $v_1$ do not show negative slope
for central collisions.
The standard JAM cascade describes well the data at 7.7 GeV,
indicating that the hadronic description may be reasonable at 7.7 GeV.
It is seen that the proton $v_1$ at 27 GeV from both JAM with attractive orbits
and standard JAM simulation yield negative slope.
Thus 7.7 and 27 GeV data do not show a hint
of the softening of the EoS within our analysis.

On the other hand,
one sees that JAM with attractive orbits
again quite reasonably describes the data at 11.5 and 19.6 GeV,
while the standard JAM cascade overestimates the data.
Therefore, STAR data on the proton directed flow
for both central and mid-central collisions
indicate evidence of the softening of the EoS.

%%%%%%%%%%%%%%%%%%%%%%%%%%%%%%%%%%%%%%%%%%%%%%%%%%%%%%%%%
\begin{figure}[tbh]
\includegraphics[width=9.0cm]{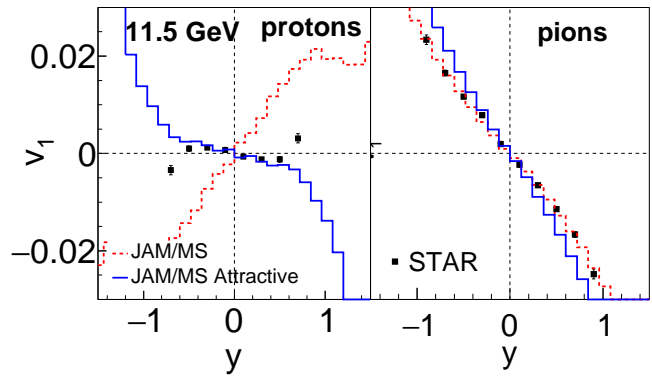}
\caption{Directed flows of protons (left) and pions (right)
in mid-central Au+Au collisions (10-40\%)
at $\srtNN=11.5$ GeV.
In the JAM calculations, momentum dependent hadronic mean-field potentials
are included (JAM/MS).
The dotted lines correspond to the result from the standard JAM/MS model,
while the solid lines are for JAM/MS with attractive orbit results.
Symbols show STAR data~\cite{STARv1}.
}
\label{fig:ms11}
\end{figure}
%%%%%%%%%%%%%%%%%%%%%%%%%%%%%%%%%%%%%%%%%%%%%%%%%%%%%%%%%

It is also necessary to examine
the influence of the nuclear mean field on the
directed flow at midrapidity, since the mean field can also modify
the flows.
Nuclear mean fields of hadrons are
included based on the framework of simplified version of 
relativistic quantum molecular dynamics (RQMD/S) in Ref.~\cite{Isse}.
Density dependent Skyrme-type and
momentum dependent Yukawa potentials are employed as in Ref.~\cite{Isse},
but with slightly different parameter sets which 
yields the incompressibility of $K=272$ MeV~\cite{qm15no}.
 
In Fig.~\ref{fig:ms11}, we show the calculated results of
the directed flow of
protons and pions at $\srtNN=11.5$ GeV
from JAM with momentum dependent potentials (JAM/MS)
together with the STAR data. 
The mean field slightly reduces the proton directed flow,
but the basic trend is the same as the JAM cascade result. 
It is interesting to see that attractive orbits
supplemented by the mean field yields negative slope,
and provides a better description of the data than the cascade calculation
at midrapidity.

\subsection{Effective EoS}
\label{sec:eos}

%%%%%%%%%%%%%%%%%%%%%%%%%%%%%%%%%%%%%%%%%%%%%%%%%%%%%%%%%
\begin{figure}[t]
\includegraphics[width=8.5cm]{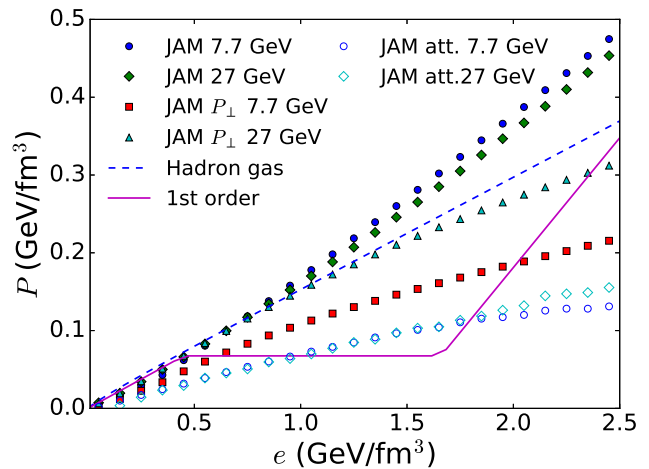}
\caption{Effective EoS extracted
from the time evolution of simulations in Au+Au collisions
at $\srtNN=7.7$ and 27 GeV.
Full (open) circles and full (open) diamonds represent
the pressures $P$ at from standard JAM (JAM with attractive orbits)
at 7.7 and 27 GeV, respectively.
Triangles and boxes represent the transverse part of the pressure $P_\perp$
for the JAM standard at 7.7 and 27 GeV, respectively.
The dashed and sold lines represent the EoS from hadron gas
and the EoS with a first-order phase transition used in Ref.~\cite{EOSQ}.
}
\label{fig:eosatt}
\end{figure}
%%%%%%%%%%%%%%%%%%%%%%%%%%%%%%%%%%%%%%%%%%%%%%%%%%%%%%%%%

We would like to see how much pressure is suppressed
by imposing attractive orbits.
The free streaming part of the local isotropic pressure $P_f$ can be computed
from the the energy-momentum tensor
$T^{\mu\nu}=\sum_h\int \frac{d^3p}{p^0}p^{\mu}p^{\nu}f_h(x,p)$ as
$P_f=-\frac{1}{3}\Delta_{\mu\nu}T^{\mu\nu}$,
with the projector of $\Delta^{\mu\nu}=g^{\mu\nu}-u^\mu u^\nu$,
where $u^\nu$ is a hydrodynamics velocity defined
by the Landau and Lifshitz definition
that may be solved iteratively~\cite{Werner:2010aa}.
The pressure difference $\Delta P$ from the free streaming $P_f$ caused by
the two-body collision between
particles $i$ and $j$ at the space-time coordinates of $q_i$ and $q_j$
is estimated based on the formula given by Ref.~\cite{Sorge:1998mk}:
\begin{equation}
  \Delta P = -\frac{\rho}{3(\delta\tau_i+\delta\tau_j)}
               (p_i'-p_i)^\mu (q_i-q_j)_\mu,
\label{eq:pre}
\end{equation}
where $\rho=N^\nu u_\nu\,,N^\nu=\sum_h\int\frac{d^3p}{p^0}p^\nu
f_h(x,p)$
is the Lorentz invariant local particle density,
$\delta\tau_{i}$ is the proper time interval between
successive collisions, and $p'_i-p_i$ is the energy-momentum change
of the particle $i$.
We extract the ``effective EoS'' by accumulating statistics
by computing local pressure $P=P_f + \Delta P$
 and energy density $e=u_\mu T^{\mu\nu}u_\nu$
at each collision point in the JAM simulation
in the central region of the reaction zone
specified by the longitudinal region $|z|<1$ fm
and the transverse radius of less than 3 fm
in Au+Au collisions at $\srtNN=7.7$, 11.5, 19.6, and 27 GeV.
We have checked that the volume dependence on the EoS
extracted from the simulation is very weak.

In Fig.~\ref{fig:eosatt}, pressure $P$ is plotted as a function
of energy density $e$ from JAM simulations as well as
the ideal hadron gas EoS and the EoS with a first-order
phase transition (EOS-Q)~\cite{EOSQ} at vanishing baryon chemical potential.
We first discuss the EoS in the standard JAM simulation.
We see some beam energy dependence of the effective EoS
as reported in Ref.~\cite{Bravina:1999dh} in which
the EoS was extracted to be $P\simeq (0.12-0.15)e$ from AGS to
Super Proton Synchrotoron (SPS) energies
that is quite similar to our results.
There is a deviation of the effective EoS from the hadron resonance gas EoS
at higher energy densities, which is mainly
due to the nonequilibrium evolution of the system
since the high energy density parts are extracted from early times
where the system is far from the equilibrium state.
In particular, pressure in the compression stage of the reaction
tends to be much higher than the values expected from the equilibrium EoS.
We note that the transverse pressure
$P_\perp=(T^{xx}_{LR}+T^{yy}_{LR})/2$, where $T^{ii}_{LR}$ is
the energy-momentum tensor at a local rest frame,
is lower than the equilibrium hadron resonance gas EoS
in high energy density regions 
as seen in our results in Fig.~\ref{fig:eosatt}
which are also reported in Ref.~\cite{Sorge:1997nv}.
It is expected that the deviation of the effective pressure in JAM compared to
the equilibrium hadron gas mainly comes from the bulk viscous pressure
and/or chemical composition.
Note that the difference between transverse and longitudinal pressure
(shear-stress tensor) should not contribute to
the isotropic pressure defined through the isotropic projection of the
energy-momentum tensor.
In the compression phase, the bulk pressure should give a positive correction
to the equilibrium  pressure, 
but in the expansion stage, the sign should change.
Judging by Fig.~\ref{fig:eosatt}, 
such sign change is, at least, not very pronounced.
So it indicates that the main contribution to the hardening of the EoS
at high energy densities in JAM is from the chemical composition;
namely, the system at early stages of the reaction is 
highly out of the chemical equilibrium state.

At lower energy densities, transverse
pressure $P_\perp$ is close to the isotropic pressure $P$, showing
the kinetic equilibration of the system, and
transverse pressure departs from isotropic pressure
above $e\approx 0.3-1.0$ GeV/fm$^3$ depending on the beam energy.
Chemical equilibrium of the system in the time evolution of hadron transport
models was investigated in Ref.~\cite{Bravina:1999dh}, and it was found that
it reaches at late times, which is consistent with the finding
here that
the effective EoS from standard JAM is close to the ideal hadron resonance
EoS at lower energy densities.
After equilibration, pressure from standard JAM approaches
values close to the ideal hadron resonance gas EoS which is seen
in Fig.~\ref{fig:eosatt}
when the energy density is less than 1 GeV/fm$^3$.
As energy density drops further, the standard JAM
yields slightly less pressure than that of the ideal gas EoS,
because of the chemical freeze-out~\cite{Sorge:1997nv}.

The effective EoS from JAM with attractive orbits is
compared with the EoS from the standard JAM simulation in Fig.~\ref{fig:eosatt}.
When attractive orbits are selected for all two-body scatterings in JAM,
we see a significant reduction of the pressure, yielding
a similar amount of softening in the transition region as EOS-Q,
although our effective EoS does not exhibit a sharp first-order
phase transition,
since we do not impose any specific condition to allow for
an attractive orbit.
We will examine which part of the EoS is responsible for the 
collapse of proton directed flow in the next sections.

\subsection{Generation of directed flow}

\begin{figure}[t]
\includegraphics[width=8.0cm]{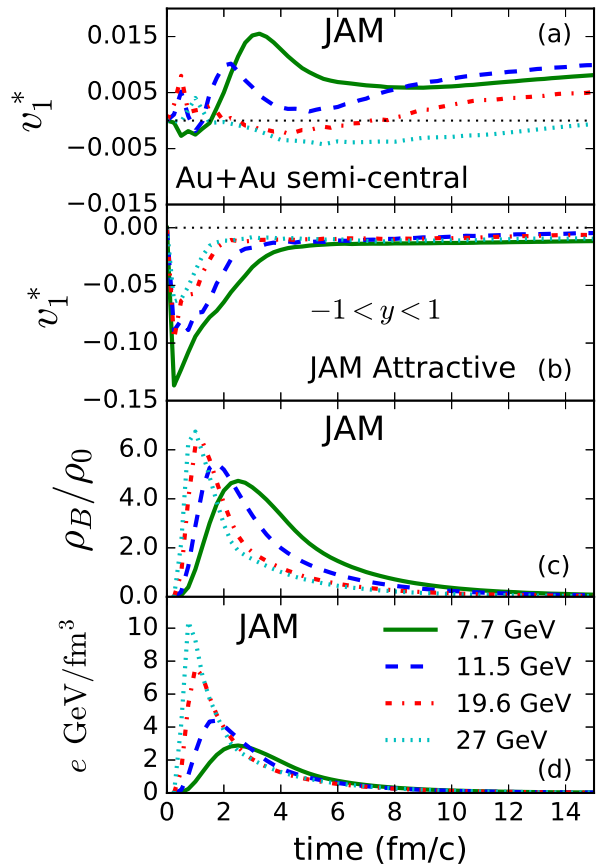}
\caption{Time evolution of sign weighted $v_1^*$
integrated over midrapidity for baryons in semicentral Au+Au collisions
for $\srtNN=7.7, 11.5, 19.6, 27$ GeV
from (a) standard JAM and (b) JAM with attractive orbits simulations.
Time evolutions of normalized net-baryon density and energy density
from the standard JAM calculations are shown
in the panels (c) and (d), respectively.
Baryon density and energy density are averaged over a cylindrical volume of
transverse radius 3 fm and longitudinal distance of 1 fm centered at the origin.
Those particles which have not interacted yet are not included
in the calculations of $v_1^*$, $\rho_B$, and $e$.
}
\label{fig:v1evol}
\end{figure}

Let us now examine where the negative slope is generated.
Time evolution of the sign weighted directed transverse momentum
integrated over the rapidity range of $-1<y<1$,
\begin{equation}
v_1^* = \int^1_{-1} dy v_1(y) \text{sgn}(y)
\end{equation}
for baryons
is displayed in Fig.~\ref{fig:v1evol} in semicentral Au+Au collision
for both the standard JAM cascade and JAM with attractive orbits.
In the standard JAM cascade, 
directed flow of baryons rises in the early states of the reaction
before two nuclei pass through each other,
and decreases with time. 
At late times, it rises slowly again with time.

%%%%%%%%%%%%%%%%%%%%%%%%%%%%%%%%%%%%%%%%%%%%%%%%%%%%%%%%%
%%%%%%%%%%%%%%%%%%%%%%%%%%%%%%%%%%%%%%%%%%%%%%%%%%%%%%%%%

In the JAM cascade with attractive orbits, on the other hand,
it is observed that directed flow is strongly modified
to be negative by the reduction of pressure at early times,
especially at lower energies, and
directed flow always increases as a function of reaction time.
We note that the slope of directed flow stays negative also in 
the rescattering stages of the reaction;
the period long after two nuclei pass through each other.
As expected, directed flow becomes smaller as collision energy becomes higher,
due to less interaction time~\cite{Csernai:2004gk}.

%%%%%%%%%%%%%%%%%%%%%%%%%%%%%%%%%%%%%%%%%%%%%%%%%%%%%%%%%
\begin{figure}[tbh]
\includegraphics[width=8.0cm]{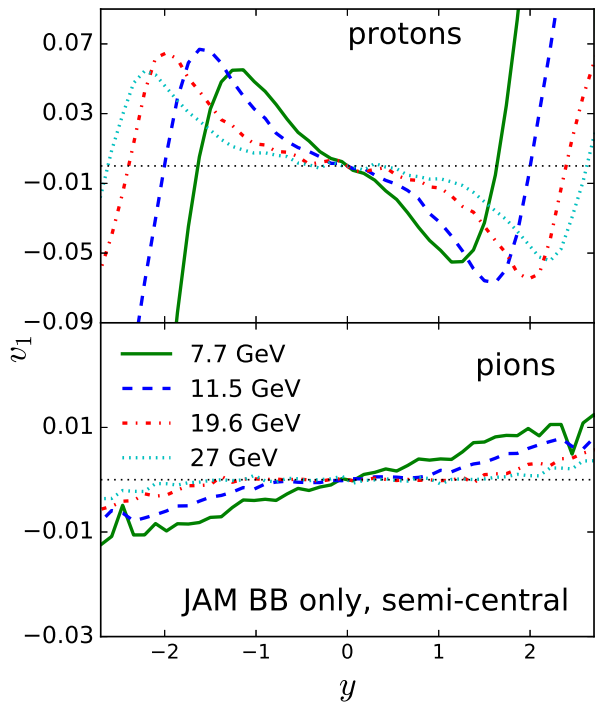}
\caption{Directed flows of protons and pions
in midcentral Au+Au collisions collisions (10-40\%).
The lines correspond to the results from the JAM cascade model
with only baryon-baryon collisions included.
}
\label{fig:v1BB}
\end{figure}
%%%%%%%%%%%%%%%%%%%%%%%%%%%%%%%%%%%%%%%%%%%%%%%%%%%%%%%%%

\subsection{Effects of meson-baryon interactions}
\label{sec:BB}

The range of beam energies covered by BES is quite interesting.
The crossing times of the two gold nuclei in the c. m. frame
are approximately 3.27, 2.15, 1.25, and 0.906 fm/$c$ at $\srtNN=7.7,
11.5, 19.6$ and 27 GeV, respectively.
Thus, hadronic rescatterings start in Au+Au collisions
before passing through two nuclei at $\srtNN=7.7$ GeV,
because crossing time is much longer than the hadronization time from
string fragmentation which is typically 1 fm/$c$.
On the other hand, at $\srtNN=27$ GeV, most of
the hadronic rescattering among produced particles occur
after two nuclei pass through each other.
In Ref.~\cite{Snellings:1999bt},
a wiggle structure in the rapidity dependence of proton directed flow
was predicted at $\srtNN=200$ GeV in which
initial nucleon-nucleon collisions are well isolated from the
late hadronic rescatterings, and  it is argued that
a wiggle structure appears 
as a result of the correlation between the position of a nucleon
and its stopping power due to initial Glauber type nucleon-nucleon collisions.
The negative directed flow seen in the hadronic transport models
at $\srtNN=27$ GeV is due to the same reason.
At energies $\srtNN \leq 19.6$ GeV,
this correlation is contaminated by meson-baryon collisions,
since  mesons and baryons start interacting with each other
before the two nuclei pass though each other.

We now look at the effects of rescatterings between mesons and baryons
in the BES energy region.
In Fig.~\ref{fig:v1BB}, we show the rapidity dependence of $v_1$
for both protons and pions for the JAM simulation without meson-baryon,
and meson-meson collisions.
We see the wiggle structure in the rapidity dependence of
the proton $v_1$ for all of the beam energies,
and pion directed flow is very small but slightly positive.
The appearance of wiggle structure
at lower beam energies when one switches off meson-baryon scatterings
may be partly due to the similar mechanism
as pointed out by Ref.~\cite{Snellings:1999bt}.
Thus we conclude that baryon-baryon collisions alone generates
negative proton directed flow at midrapidity, and meson-baryon collisions
bring the proton directed flow to the positive side
and the pion directed flow to the negative side.
This implies that strong negative directed flow must be generated at the
initial stages of nuclear collisions  to get the negative proton
flow at freeze-out.

%%%%%%%%%%%%%%%%%%%%%%%%%%%%%%%%%%%%%%%%%%%%%%%%%%%%%%%%%
\begin{figure}[t]
\includegraphics[width=8.0cm]{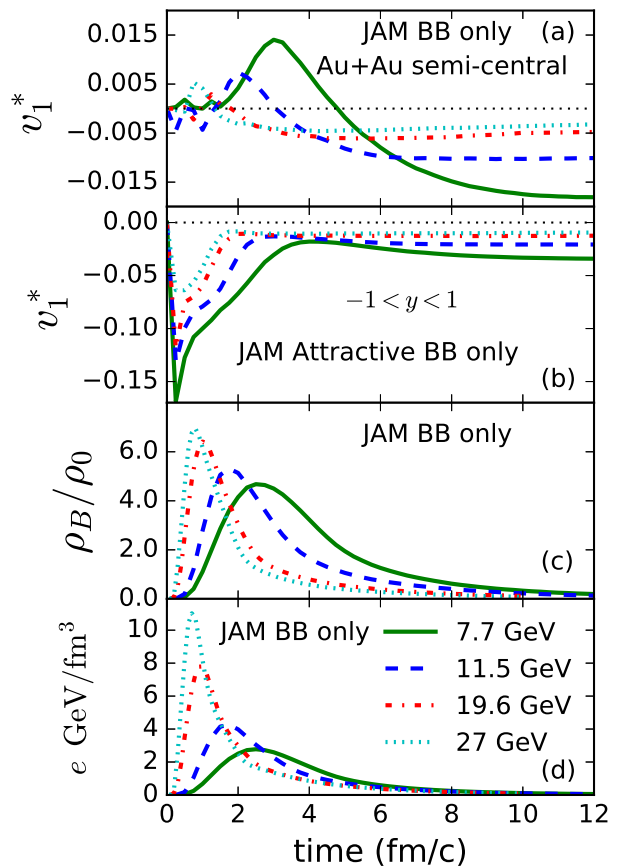}
\caption{
Same as Fig.~\ref{fig:v1evol}, but JAM cascade simulations with 
only baryon-baryon collisions.
}
\label{fig:v1evolBB}
\end{figure}
%%%%%%%%%%%%%%%%%%%%%%%%%%%%%%%%%%%%%%%%%%%%%%%%%%%%%%%%%

In order to examine further in detail how $v_1$ is generated,
we plot in Fig.~\ref{fig:v1evolBB} the time evolution of
$v_1^*(|y|<1)$ from the JAM simulations
by switching off all baryon-meson and meson-meson collisions (JAM BB only).
The upper panel of Fig.~\ref{fig:v1evolBB} shows the time evolutions
of $v_1^*$ from JAM simulations with only baryon-baryon collisions.
It shows that positive directed flow is first generated
before two nuclei pass though each other,
and then it becomes negative in the expansion stages of the reaction.
The lower beam energy yields larger negative directed flow.
This is because of the effect of spectators as well as
the increasing number of baryon-baryon collisions at lower beam energies.
The average number of baryon-baryon collisions $\langle N_\text{coll}\rangle$
in Au+Au semicentral collisions
is larger at lower beam energy: 
$\langle N_\text{coll}\rangle=490, 400, 340$, and 310
for $\srtNN=7.7, 11.5, 19.6$ and 27 GeV, respectively.

In the middle panel of Fig.~\ref{fig:v1evolBB},
we display the time evolution of $v_1^*$
in JAM BB only with attractive orbits.
As in the case of full simulation, strong negative
directed flow is generated in the early stages of the reaction
and it rises in time, then it stays the same value at later times
because of the absence of hadronic rescatterings.
The question then arises, which part of the reaction stage is more
relevant for the negative directed flow of protons ?

%%%%%%%%%%%%%%%%%%%%%%%%%%%%%%%%%%%%%%%%%%%%%%%%%%%%%%%%%
\begin{figure}[tbh]
\includegraphics[width=8.5cm]{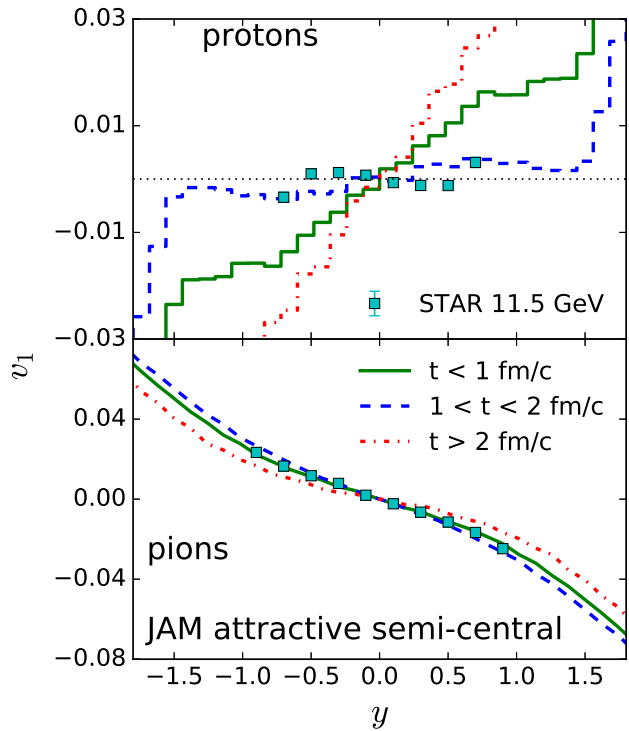}
\caption{Rapidity distribution of directed flow $v_1$ 
for protons and pions in semicentral Au+Au collisions
for $\srtNN=11.5$ GeV from JAM simulation in which attractive orbits
are imposed at different time intervals;
the solid line corresponds to the simulation with attractive
orbits for time $t< 1$ fm/$c$, 
the dashed line for $1< t< 2$ fm/$c$, 
and
the dotted line for $t > 2$ fm/$c$.
}
\label{fig:v1tcut}
\end{figure}
%%%%%%%%%%%%%%%%%%%%%%%%%%%%%%%%%%%%%%%%%%%%%%%%%%%%%%%%%

\begin{figure}[tbh]
\includegraphics[width=8.5cm]{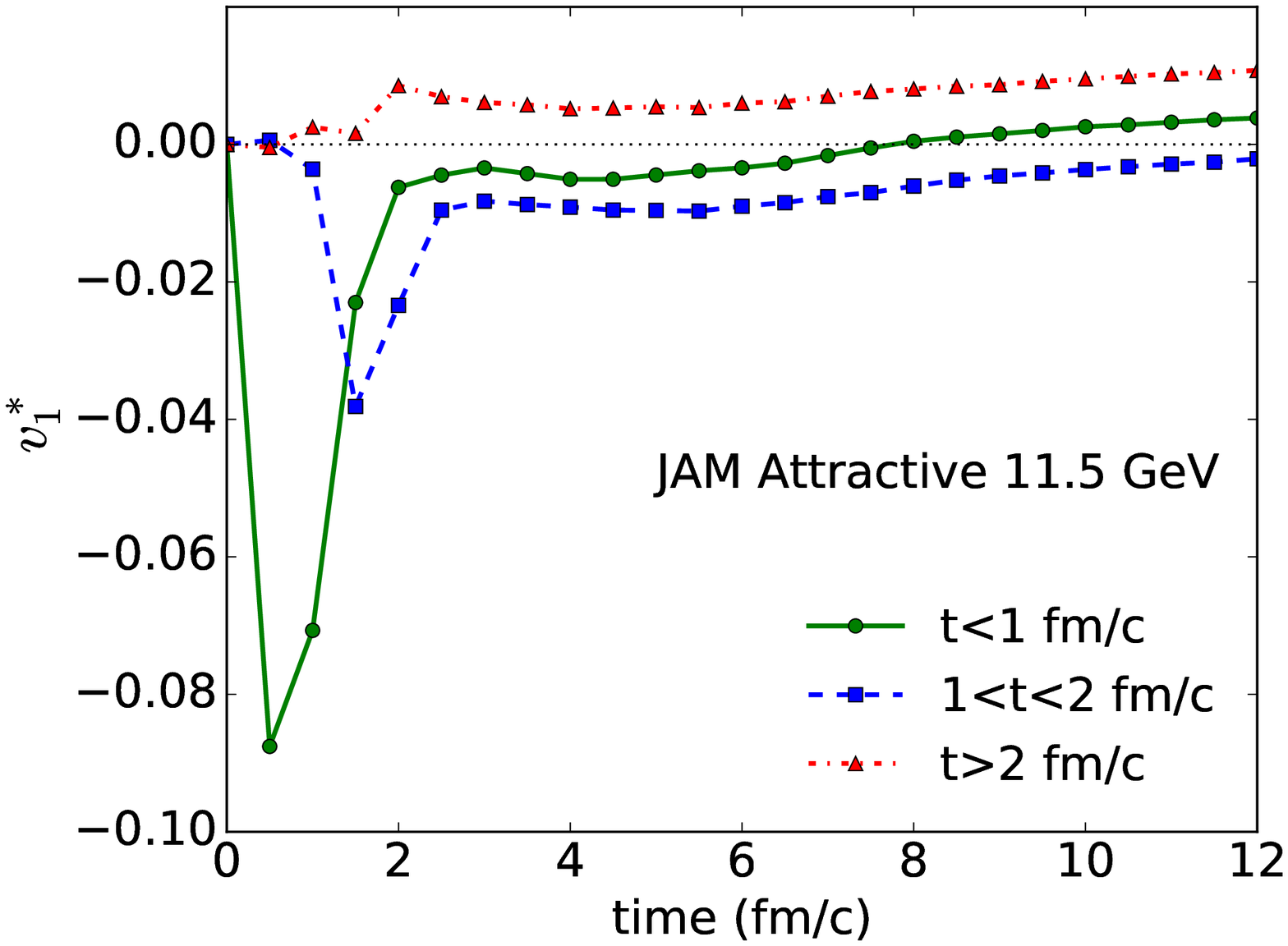}
\includegraphics[width=8.0cm]{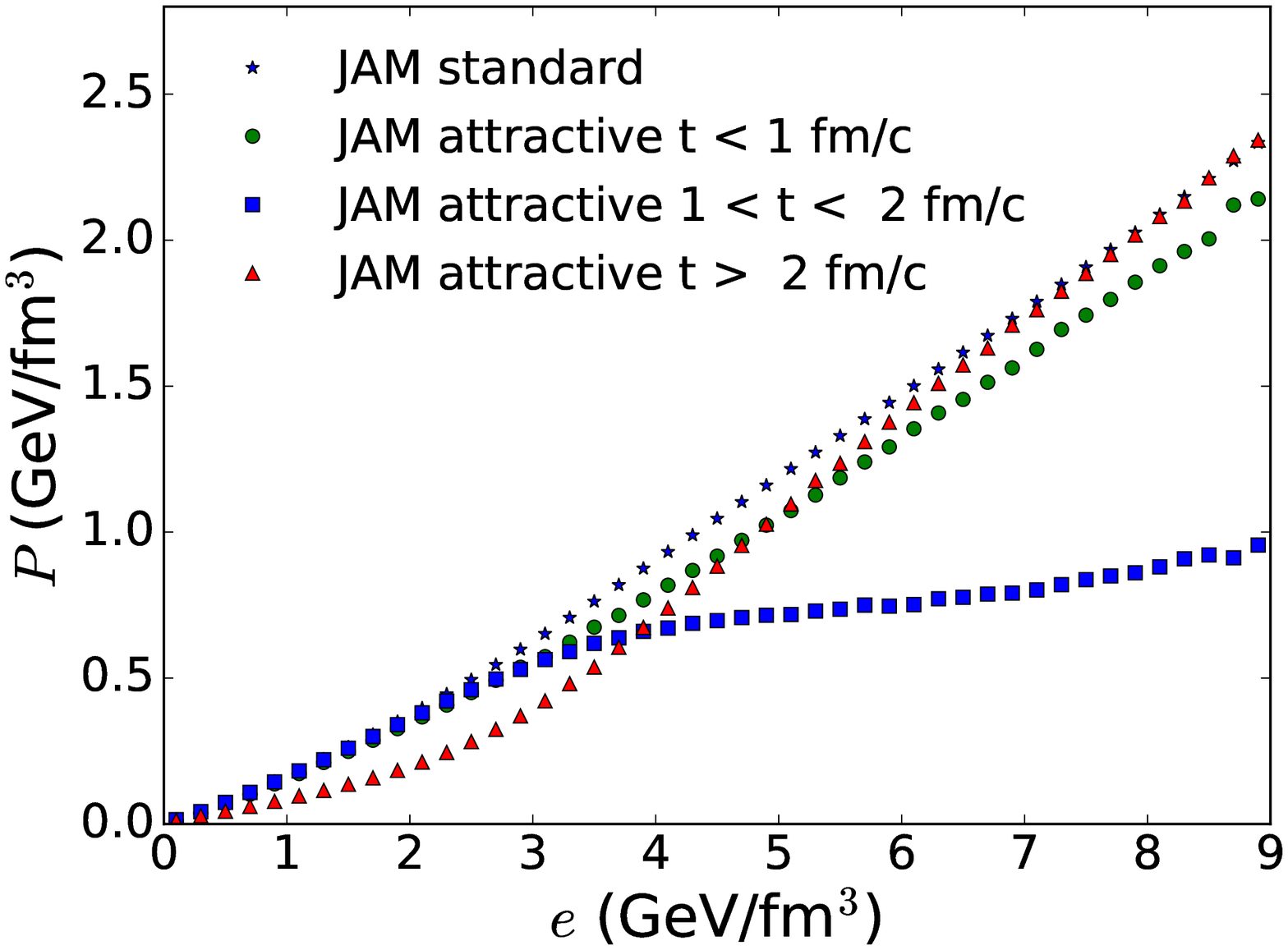}
\caption{Time evolution of sign weighted $v_1^*$ for
baryons (upper panel) and effective EoS (lower panel)
in Au+Au collision at $\srtNN=11.5$ GeV
from JAM simulations in which attractive orbits
are imposed at different time intervals;
the circles correspond to the simulation with attractive
orbits for time $t< 1$ fm/$c$, 
the squares for $1< t< 2$ fm/$c$, 
and
the triangles for $t > 2$ fm/$c$.
}
\label{fig:v1tcuteos}
\end{figure}

To try to answer the question,
we plot in Fig.~\ref{fig:v1tcut}
the rapidity dependence of $v_1$ in Au+Au semicentral collision
for three different JAM simulations at $\srtNN=11.5$ GeV
by noting that the crossing time of two nuclei is about 2.15 fm/$c$:
(1) JAM with attractive orbits only for compression stages of
the reaction until two nuclei reach full overlap,
i.e. time less than 1 fm/$c$;
(2) JAM with attractive orbits at the reaction time between 1 and 2 fm/$c$
which corresponds to the time range with largest baryon densities.
(3) JAM with attractive orbits at times later than 2 fm/$c$; and
As we expect, when attractive orbits are imposed at times
later than 2 fm/$c$, it is too late to generate negative flow,
and the result is almost identical to the standard JAM simulation.
Furthermore, it is very interesting to see that
the effect of attractive orbits at times earlier than 1 fm/$c$
is small and its effect alone cannot explain the strong suppression of the flow.
To see this point more clearly, time evolution of $v_1^*$
is displayed in Fig.~\ref{fig:v1tcuteos}.
One see that strong negative directed flow generated in the earliest stages
of the reaction quickly disappears with a much faster rate than the one
shown in the lower panel of Fig.~\ref{fig:v1evol}.
Thus initial scatterings at the compression stage of the reaction
are not important in generating negative $v_1$.
Finally, it is shown that what is responsible for the strong suppression
of the proton flow is the effect of attractive orbits in the time interval
$1< t < 2$ fm/$c$, which coincides with the highest baryon
density in the course of the reaction at $\srtNN=11.5$ GeV.
This effect can be further confirmed in the time evolution
of $v_1^*$ in Fig.~\ref{fig:v1tcuteos}.
Attractive orbits at this stage are important to suppress the rise of $v_1$
due to the hadronic rescatterings.

%%%%%%%%%%%%%%%%%%%%%%%%%%%%%%%%%%%%%%%%%%%%%%%%%%%%%%%%%
%%%%%%%%%%%%%%%%%%%%%%%%%%%%%%%%%%%%%%%%%%%%%%%%%%%%%%%%%

Effective EoS for each simulation are plotted together with the
EoS of the standard JAM simulation
in the lower panel of Fig.~\ref{fig:v1tcuteos}.
The effect of the attractive orbits for $t<1$ fm/$c$ is the slight reduction
of pressure at high energy densities; on the other hand,
attractive orbits for $t>2$ fm/$c$ reduce the pressure at lower energy
densities.  Attractive orbits at $1<t<2$ fm/$c$ strongly reduces
the pressure at high energy densities which results in the
collapse of proton directed flow.
However,  it does not necessarily imply
that the equilibrium  EoS at high energy density at high baryon density 
needs to be very soft for the negative $v_1$,
since most of pressures at energy densities above 4 GeV/fm$^3$
are extracted from preequilibrium stages of the reaction in the JAM simulation.
Nonequilibrium effects need to be examined.
Nevertheless, it suggests the need of a nonstandard dynamical effect
which is related to the reduction of pressure at high baryon densities.

This analysis strongly suggests the importance of reaction dynamics
at high baryon density. 
The time period of $1<t<2$ fm/c at $\srtNN=11.5$ GeV
relevant to the negative flow
corresponds to the preequilibrium stage
in the JAM hadronic transport approach,
even though produced hadrons start to interact with each other.
We do not have partonic interactions,
unlike the PHSD model~\cite{Konchakovski:2014gda}.
The effects of partonic interactions in the early stages of
the reaction should be examined elsewhere
in order to understand the dynamical effects on the directed flow.

\subsection{Elliptic flow}

\begin{figure}[tbh]
\includegraphics[width=8.5cm]{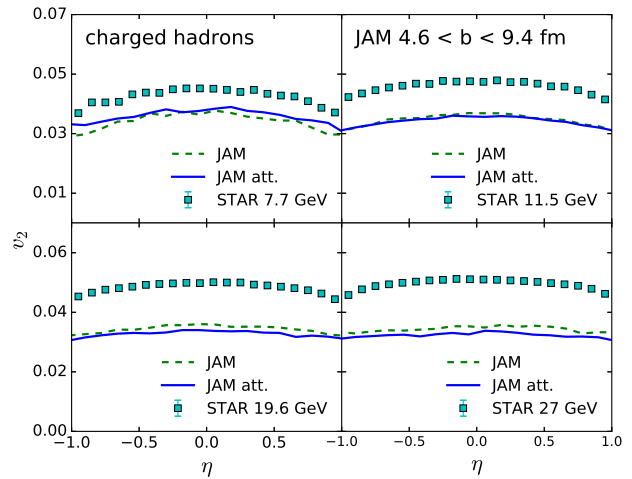}
\caption{Elliptic flows of charged hadrons
in midcentral Au+Au collisions (10-40\%).
The dotted lines correspond to the result from the standard JAM cascade model,
while the solid lines are for the JAM cascade with attractive orbit results.
Symbols show STAR data~\cite{STARv2}.
}
\label{fig:v2ch}
\end{figure}

In order to see systematics on the use of the attractive orbits
in the scattering style, it is important to check
other flow harmonics such as elliptic flow
$v_2=\langle \cos(2\phi)\rangle$.
In Fig.~\ref{fig:v2ch}, JAM cascade results are
compared with the pseudorapidity dependence of $v_2$
for charged hadrons in midcentral (10-40\%) Au+Au collisions
at $\srtNN=7.7, 11.5, 19.6$ and $27$ GeV~\cite{STARv2}.
It is seen that $v_2$ at midrapidity is not modified
by the attractive orbits scattering style, 
but it underestimates the data about 20--30\%.
Therefore, we do not see any softening effects on $v_2$ within our approach.
Underestimation of $v_2$ by our approach suggests the need of
partonic interactions in the early stages of the reactions.
We have also studied other inclusive hadronic observables such
as transverse momentum distributions and rapidity distributions
and found that the effect of attractive orbits on them is very small.

%%%%%%%%%%%%%%%%%%%%%%%%%%%%%%%%%%%%%%%%%%%%%%%%%%%%%%%%%
%%%%%%%%%%%%%%%%%%%%%%%%%%%%%%%%%%%%%%%%%%%%%%%%%%%%%%%%%

It is also important to examine other observables
such as the net-baryon number cumulants
in the same energy range~\cite{Adamczyk:2013dal}.
If the softening of the EoS comes from criticality around the critical point,
divergence of cumulants appears as oscillating behavior
as a result of smearing by the finite quark mass~\cite{ScalingFn}
or finite volume~\cite{IMO2015}.
Thus it is an interesting question whether dynamical model calculations 
with the EoS softening can describe the observed nonmonotonic behavior
of cumulant ratios~\cite{Adamczyk:2013dal}.
Recently, it was shown that JAM with attractive orbits
as well as nuclear mean-field effects does not describe
the observed large enhancement of cumulant ratios~\cite{He:2016uei}.

\subsection{EoS dependence of directed flow}
\label{sec:eosdep}

\begin{figure}[thb]
\includegraphics[width=8.5cm]{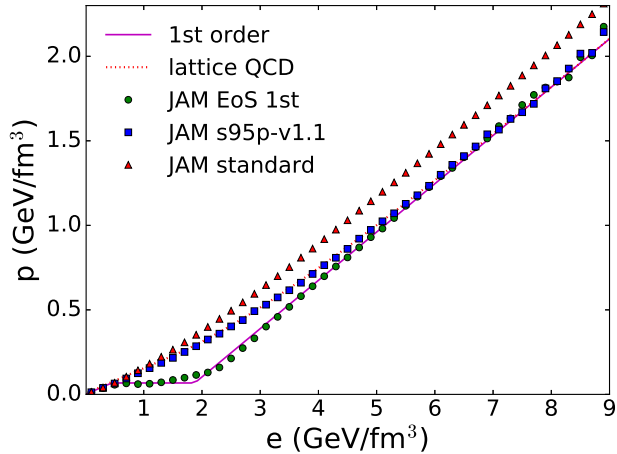}
\caption{Equation of state implemented into JAM simulations.
The sold line represents the first-order phase transition EoS, and
the dotted line represents the EoS from lattice QCD~\cite{Huovinen:2009yb}.
The circles and squares are results obtained by the JAM simulations,
which are almost identical to the inputs.
Effective EoS obtained from JAM standard simulation is also displayed by
the triangles.
}
\label{fig:jameos}
\end{figure}

So far, we implement attractive orbits in JAM for all hadron-hadron
$2\to2$ scatterings without any restrictions.
As a result, our equation of state is soft for all energy densities
as shown in Fig.~\ref{fig:eosatt}.
We now explore the EoS dependence of the directed flow. For this purpose,
instead of imposing attractive scattering all the way,
we select attractive orbits at each two-body scattering
with the probability $p_\text{attractive}$
given by 
\begin{equation}
p_\text{attractive}= \max\left(0,\frac{P_f - P(e)}{P(e)}\right),
\end{equation}
where $P(e)$ is a pressure as a function of energy density $e$ from a given
EoS, and $P_f$ is the local pressure at the collision point computed from
the energy-momentum tensor as in Sec.~\ref{sec:eos}.
The QCD equation of state at high baryon densities is not well understood.
As a first step, we use the EoS which does not depend on
baryon density for simplicity, since our purpose here is to check
the systematics of our approach that modifies the scattering style.
We test the EoS with crossover from lattice QCD ($s95p$-v1.1)
taken from Ref.~\cite{Huovinen:2009yb,Hirano:2012kj}, 
and the EoS with a first-order phase transition
similar to that of EOS-Q with the modification of
the slope in the QGP phase to $p=e/3.5$
(instead of the massless ideal gas EoS $p=e/3$)
so that pressure at high energy densities
is consistent with the lattice EoS as shown in Fig.~\ref{fig:jameos}.
We also show in Fig.~\ref{fig:jameos}
the results which are extracted from JAM simulations
to ensure that our simple approach is consistent with a given EoS.
It is seen that our simple approach works very well to modify the EoS
of the system.

%%%%%%%%%%%%%%%%%%%%%%%%%%%%%%%%%%%%%%%%%%%%%%%%%%%%%%%%%
%%%%%%%%%%%%%%%%%%%%%%%%%%%%%%%%%%%%%%%%%%%%%%%%%%%%%%%%%

%%%%%%%%%%%%%%%%%%%%%%%%%%%%%%%%%%%%%%%%%%%%%%%%%%%%%%%%%
\begin{figure}[tbh]
 \includegraphics[width=8.5cm]{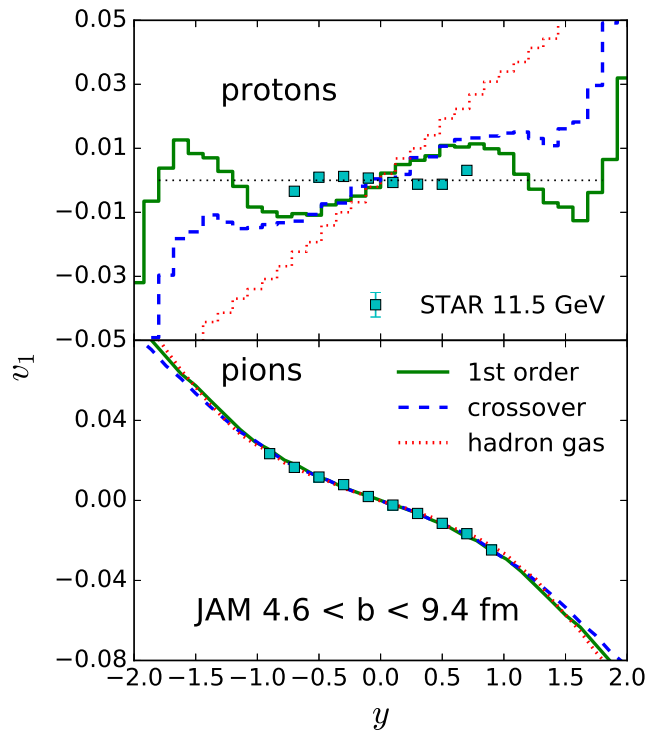} 
\caption{Rapidity distribution of directed flow $v_1$ for protons and pions
in semicentral Au+Au collisions
for $\srtNN=11.5$ GeV from JAM simulations
with different EoS.  The solid line presents the JAM result
with the EoS with fist-order phase transition,
and the dashed line presents the JAM result with the
crossover EoS.  The dotted line is for the standard JAM result.}
\label{fig:v1eos} \end{figure}
%%%%%%%%%%%%%%%%%%%%%%%%%%%%%%%%%%%%%%%%%%%%%%%%%%%%%%%%%

In Fig.~\ref{fig:v1eos}, we plot the rapidity dependence of
the directed flow of protons (upper panel) and pions (lower panel)
in Au+Au semicentral collision at $\srtNN=11.5$ GeV
obtained from the EoS with crossover and first-order phase transition.
While pion flow is not sensitive to the EoS,
one sees that both EoSs yield the suppression of proton directed flow
compared to the standard cascade simulation, 
since both EoSs are softer than the effective EoS
from the standard JAM cascade simulation
as compared in Fig.~\ref{fig:v1eos}.
Proton flow at midrapidity $|y|<1.0$ is not sensitive to the EoS,
but two EoS give different behavior for larger rapidities indicating that
the softening point of the current EoS is responsible to
the rapidity $|y|>1.0$ at $\srtNN=11.5$ GeV, and explicit EoS dependence
of the directed flow may be observable in the experiments.
It remains for further work to establish the EoS dependence of the
directed flow by utilizing a fully baryon density dependent EoS.
The interactions used here do not employ any baryon density dependent
interactions, $U(\rho)$, as one may want to
use in relativistic mean-field models of high baryon density matter.
It is also possible that  $\Delta$ matter with slightly nonuniversal scalar
attraction can easily cause a first-order phase transition
without mentioning any high baryon density QCD.
We will present a detailed systematic study of the EoS dependence of
the directed flow elsewhere.

\section{Summary}

In summary,
we have investigated the effect of the softening of the EoS
on the directed flow of protons and pions within a microscopic
transport approach.
The transport model JAM with standard stochastic two-body
scattering style predicts
the large positive slope of proton $v_1$ at collision energy
below $\srtNN=19.6$ GeV, and the negative slope of proton $v_1$
only at higher collision energy $\srtNN \geq 27$ GeV,
which disagree with the STAR data.
However,
softening effects of the EoS simulated by attractive orbit scatterings
lead to a dramatic change in the dynamics,
and yield significant reduction of proton $v_1$ which
well describe the STAR data around the minimum of $dv_1/dy$
at $10 \lesssim\srtNN\lesssim 20$ GeV.
The softening effects were not needed in the present approach
at lower energies, $\srtNN=7.7~\GeV$.
We found that attractive orbit scattering style
does not modify elliptic flow at mid-rapidity.
We show that this softening effect is needed only at early stages 
of the reaction where the system reaches the high baryon density
state at midrapidity.

We also proposed a simple recipe to simulate a given EoS within a
hadronic transport model, and compared two different EoS.
We saw an EoS dependence of
the proton directed flow in the forward rapidities.
More detailed systematic studies are needed,
using a fully baryon density dependent EoS,
in order to draw  a conclusion that the minimum of $dv_1/dy$
is a result of the softening of the EoS
which may be caused by a first-order phase transition
~\cite{Stoecker:2004qu,Brachmann:1999xt,Gyulassy:1981nq,Sorge:1998mk}.

A possible scenario to fully explain
the beam energy dependence of the directed flow
may be described as follows.
We assume that there exists the softest point
in the energy density range reachable at $\srtNN=11.5~\GeV$.
Hadrons will feel an attractive force
when they go across the surface of the soft region.
This additional force can be simulated by introducing
the attractive orbit scatterings among hadrons, 
as we have demonstrated in the present work,
and negative $dv_1/dy$ emerges.
One may need to introduce new degrees of freedom other than hadrons
to understand the rehardening at higher energies.

It seems obvious to infer a softening of the EoS
from the experimentally observed collapse of
 net-proton flow when the c.m. energy is increased from 7 to 11 GeV.
However, the statement of a discovery of the ``softening'' of the EoS from the
net-proton $v_1$ data shows even more convincing evidence
for the ``phase transition'' as we observe the rebound at higher energies;
namely the STAR-observed second change of sign of the $v_1$
values of the net protons at $\srtNN \approx 40$ GeV
back to positive $v_1$ at higher energies~\cite{STARv1}.
This shows that the soft region is overcome, 
and the directed flow picks up steam again, due to the rehardening
of the EoS at considerably larger energy densities.

In the near future, a more detailed analysis of the
softening effect should be addressed by employing a realistic
EoS which is consistent with the lattice QCD result.
Because of the nonequilibrium evolution,
the pressure generation due to the two-body collision
$\Delta P$ depends in the current study not only on the difference between the
equilibrium EoSs, but also on the dynamical evolution of the system.
Perhaps a more justified way of fixing the EoS would be to look at a fully
equilibrated system in a box, and then determine the probabilities
for attractive orbits as a function of the energy of the colliding particles,
so that a given (equilibrium) EoS would be reproduced;
then the microscopic dynamics would tell how the system looks
in out-of-equilibrium situations. 

It is expected that properties of the EoS at high
baryon density may be probed sensitively by using the flow.
Future experiments such as the BES II at RHIC~\cite{BESII},
FAIR~\cite{FAIR},
NICA~\cite{NICA},
and J-PARC~\cite{Sako:2014fha}
should clarify this point at lower collision energies $\srtNN\leq10$ GeV.

\section*{Acknowledgement}
We would like to thank Adrian Dumitru for valuable comments.
H.S. thanks Nu Xu, Zangbu Xu, Declan Keane, and Paul Sorensen
for numerous useful discussions.
Y.N. thanks the Frankfurt Institute of Advanced Studies where part of this
work was done.
%Y. N. is grateful to Harri Niemi for illuminating discussions.
This work was supported in part by 
the Grants-in-Aid for Scientific Research from JSPS
%from the Japan Society for the Promotion of Science (JSPS)
(No.
 15K05079%, %AO & YN (Kiban C)
,
 No. 15H03663 %((B) PI:A.Nakamura) )
and
 No. 15K05098 %((C TK & YN ) )
),
the Grants-in-Aid for Scientific Research on Innovative Areas from MEXT
 (No. 24105001 and No. 24105008),
and
by the Yukawa International Program for Quark-Hadron Sciences.
H.N. has received funding from the European Union's Horizon 2020 Research and
Innovation Programme under Marie Sklodowska-Curie Grant Agreement
No. 655285 and from the Helmholtz International Center for FAIR
within the framework of the LOEWE program launched by the State of Hesse.

\end{document}